\documentclass[12pt]{article}%
\usepackage{amsmath}
\usepackage{amsfonts}
\usepackage{amssymb}
\usepackage{graphicx}
\usepackage{epsf,color,colordvi,pifont}
\usepackage{amsfonts}%
\providecommand{\U}[1]{\protect\rule{.1in}{.1in}}
\begin{document}

\title{Classical Noncommutative Electrodynamics with External Source}
\author{{\large T. C. Adorno}\thanks{adorno@dfn.if.usp.br}{\large ,} {\large D. M.
Gitman}\thanks{gitman@dfn.if.usp.br}\\\textit{Instituto de F\'{\i}sica, Universidade de S\~{a}o Paulo,}\\\textit{Caixa Postal 66318, CEP 05508-090, S\~{a}o Paulo, S.P., Brazil}\\{\large A.~E.~Shabad}\thanks{shabad@lpi.ru}\\\textit{P.~N.~Lebedev Physics Institute, Moscow, Russia}\\{\large D. V. Vassilevich}\thanks{dvassil@gmail.com}\\\textit{CMCC - Universidade Federal do ABC, Santo Andr\'e, S.P., Brazil}\\\textit{Department of Physics, St.~Petersburg State University, Russia} }
\maketitle

\begin{abstract}
In a $U(1)_{\star}$-noncommutative (NC) gauge field theory we extend the
Seiberg-Witten (SW) map to include the (gauge-invariance-violating) external
current and formulate -- to the first order in the NC parameter --
gauge-covariant classical field equations. We find solutions to these
equations in the vacuum and in an external magnetic field, when the 4-current
is a static electric charge of a finite size $a,$ restricted from below by the
elementary length. We impose extra boundary conditions, which we use to rule
out all singularities, $1/r$ included, from the solutions. The static charge
proves to be a magnetic dipole, with its magnetic moment being inversely
proportional to its size $a$. The external magnetic field modifies the
long-range Coulomb field and some electromagnetic form-factors.

We also analyze the ambiguity in the SW map and show that at least to the
order studied here it is equivalent to the ambiguity of adding a homogeneous
solution to the current-conservation equation.

\end{abstract}

\section{Introduction}

Noncommutative (NC) field theories, based on a profound revision of the most
fundamental properties of the space-time achieved by introducing an elementary
length, play a challenging role in modern theoretical physics. These theories
do not need a lengthy introduction, we refer the interested reader to the
review papers \cite{DouNek01,Szabo}. The present paper is devoted to a
construction of an NC extension of electrodynamics in its most classical
sense. It is notable that the resulting electrodynamics is, already at the
classical level, a nonlinear theory, rich in properties. For instance, it
possesses the birefringence and photon splitting in external fields
\cite{FreGitSha11}. But, in contrast to other nonlinear theories,
\textit{e.g.,} the classical Yang-Mills theory, or the Born-Infeld
electrodynamics, or else Quantum Electrodynamics (QED) after radiative
corrections in it are taken into account, the classical NC electrodynamics is,
besides, intrinsically anisotropic. We shall demonstrate below that the
classical NC electrodynamics reproduces also other interesting features, known
\cite{BerLifPit}, \cite{ShaUso10} in QED. Moreover, we establish that
electrostatic charge with its density homogeneously distributed over a
finite-size sphere carries a magnetic moment depending on its size. Hence, an
idea of the NC magnetic moment of the proton appears. In an external magnetic
field we find a modification of the Coulomb law at large distances from the
charge of a sort completely alien to QED. This effect may be referred to as a
macroscopic manifestation of the elementary length at large distances. A study
of these and the like classical phenomena is inevitable, as long as possible
observable consequences of noncommutativity are to be looked for.

In the present paper we do not quantize the electromagnetic field. The charge
carriers are represented through the currents rather than through elementary
fields. This latter task, namely the introduction of currents, will appear to
be rather nontrivial. It is very well known that there are severe restrictions
on the gauge groups and their representations so that the gauge
transformations might form a closed algebra on the NC plane \cite{gauge}. To
overcome this difficulty one either uses the Seiberg-Witten (SW) map
\cite{SeiWit99}, or makes the gauge transformations twisted
\cite{twist1,twist2}. None of these are strictly speaking necessary in NC
electrodynamics, since the $U(1)$ gauge group can be nicely deformed to a
$U(1)_{\star}$ group. Therefore, many papers define noncommutative
electrodynamics as a $U(1)_{\star}$ gauge theory\footnote{There is another,
less frequently used terminology, see \cite{AdoBalGit10}, according to which
this deformation is called the Moyal modification.}, see, \textit{e.g.},
\cite{ChaJab01,Hayakawa00,RiaJab00,ArdSad01}. Nevertheless, various aspects of
the SW map were developed for the NC $U(1)$ theories
\cite{AsaKis99,BicGriPopSchWul01}. Since the electromagnetic potential after
the SW map has the standard gauge transformation properties, this map
facilitates the analysis of phenomenological predictions of NC theories
\cite{CarHarKosLanOka01,Cai01,GurJacPiPol01,Chatillon:2006rn}. Also, the SW
map has interesting effects on renormalizability of NC field theories even in
the $U(1)$ case
\cite{BicGriPopSchWul01,Wulkenhaar02,FruGriMorPopSch02,BirLatRadTra10}.

In this work we study, to the lowest order of noncommutativity, a NC Maxwell
theory in the presence of sources. Clearly, if the phenomenological analysis
involves a comparison of solutions in a commutative theory to noncommutative
corrections, it is essential that both commutative and noncommutative fields
have the same transformation properties under the gauge group. For example,
the electric and magnetic fields must be gauge invariant. In other words, for
such applications one has to introduce commutative fields into a
noncommutative field theory\footnote{
%We like to s
Stress, that for some applications, like the recent analysis of noncommutative
Dirac quantization condition \cite{ngvik:2011gb}, this is not necessary.}.
This is precisely what the SW map does.

In Section \ref{sec-Max} , as a preparation to the SW construction, we first
study the $U(1)_{\star}$ gauge theory with currents. We observe, that although
the set of equations of motion consisting of the Maxwell equation and the
current conservation condition is gauge covariant, the action is not gauge
invariant. This fact is analogous to a well-known property of nonabelian
(``commutative'') Yang-Mills theories and does not mean by itself any internal
inconsistency. However, for the SW map it implies that this map has to be
performed in equations of motion rather than at the level of the action. We
proceed in this way and rederive the SW map for currents \cite{Ban} in the
first order of the noncommutativity parameter. Field equations include
potentials along with the field strengths. Their gauge covariance is
effectuated via the statement that the gauge-transformed potentials satisfy
the same equations within the accuracy adopted in the paper. Moreover,
solutions to the equations, from which all potential-containing terms are
dropped, satisfy the original equations with the potentials retained.

In Section \ref{sec-mag} we consider first-order NC corrections to the field
of a static spherically symmetric charge distributed over a sphere of finite
size, assuming that the noncommutativity is space-space. It is important that
the size of the charge be larger than the elementary length characteristic of
the NC theory. In contrast to our previous paper \cite{AdoGitShaVas11}, we
place the static delocalized charge in a constant and homogeneous external
magnetic field. This gives rise to NC corrections to the electrostatic
potential, linear with respect to the charge and to the external magnetic
field, as well as to appearance of NC magnetic field, produced by the charge,
quadratic with respect to its value and independent of the external magnetic
field. In Subsection \ref{subsection3.1} we impose boundary conditions onto
the field equations that would exclude a singular behavior of their solutions
in the origin (where the charge is centered) and find magnetic and
electrostatic field, produced by the static charge. In Subsections
\ref{subsection3.2}, \ref{subsection3.3} we consider other magnetic and
electric solutions that are either singular in the origin or do nor decrease
in the remote region and discuss which solutions should be selected as
physical and associated with the charge. It is notable that the physical
(regular in the origin) magnetic solution does not have a finite limit if the
size of the charge is taken infinitely small. It is not obliged to, indeed,
since no size of any physical object may be smaller than the elementary
length. In Section \ref{sec-form} we discuss various peculiarities carried by
the solutions found: the magneto-electric effect and, especially, the NC
magnetic moment intrinsic to a charged extended particle, which is inversely
proportional to its size and, hence, most important \cite{AdoGitShaVas11} for
particles that are considered point-like within the present experimental
possibilities, like charged leptons and quarks. Also the corresponding
hyperfine splitting caused by this moment if the particle is taken as an
atomic nucleus (Subsection \ref{subsection4.1}) is considered. In Subsection
\ref{subsection4.2} the NC correction to the electrostatic potential is
inspected that consists of an anisotropic Coulomb field and of an electric
quadrupole contribution. An NC analog of the Zeeman splitting is pointed. In
Subsection \ref{subsection4.3} we propose an extension of the Furry theorem in
application to NC electrodynamics that explains, on general basis, the
character of the dependence of the magnetic and electrostatic solutions upon
powers of the charge, external field and the NC parameter.

It is known, that the SW map is not unique. In Section \ref{sec-amb} we show
that -- to the first order at least -- the ambiguity in the SW map for the
current derived in Section \ref{sec-Max} is precisely the ambiguity of adding
a homogeneous solution of the charge-conservation equation.

\section{Maxwell equations on noncommutative space-time}

\label{sec-Max}

\subsection{$U(1)_{\star}$ gauge theory}

In this paper we work on the Moyal plane, which is (identified with) a space
of sufficiently smooth functions on $\mathbb{R}^{4}$ equipped with the Moyal
star product
\begin{equation}
f\left(  x\right)  \star g\left(  x\right)  =f\left(  x\right)  e^{\frac{i}%
{2}\overleftarrow{\partial}_{\mu}\theta^{\mu\nu}\overrightarrow{\partial}%
_{\nu}}g\left(  x\right)  \, \label{Mstar}%
\end{equation}
where the NC parameter $\theta^{\mu\nu}$ is assumed to be constant and small.

We start with the action of a NC $U(1)_{\star}$ gauge theory
\begin{align}
&  \check{S}=\check{S}_{\mathrm{A}}+\check{S}_{\mathrm{jA}}\,,\label{SAjA}\\
&  \check{S}_{\mathrm{A}}=-\frac{1}{16\pi c}\int dx\check{F}_{\mu\nu}%
\star\check{F}^{\mu\nu}\,,\quad\check{S}_{\mathrm{jA}}=-\frac{1}{c^{2}}\int
dx\check{j}^{\mu}\star\check{A}_{\mu}\,,\nonumber\\
&  \check{F}_{\mu\nu}=\partial_{\mu}\check{A}_{\nu}-\partial_{\nu}\check
{A}_{\mu}+ig\left[  \check{A}_{\mu}\overset{\star}{,}\check{A}_{\nu}\right]
\,,\qquad\left[  \check{A}_{\mu}\overset{\star}{,}\check{A}_{\nu}\right]
=\check{A}_{\mu}\star\check{A}_{\nu}-\check{A}_{\nu}\star\check{A}_{\mu
}\,,\nonumber
\end{align}
where the coupling constant is $g=e/(\hbar c),$ with $e$ being the elementary
charge (for the electron $e=-\left\vert e\right\vert $). The action $\check
{S}_{\mathrm{jA}}$ may be obtained from an NC theory with fundamental fields.
For example, for the case of fundamental fermions we have
\begin{equation}
\check{S}_{\check{\psi}}=i\int dx\,\overline{\check{\psi}}\star\gamma^{\mu
}\left(  \partial_{\mu}-ig\check{A}_{\mu}\star\right)  \check{\psi}\equiv
i\int dx\overline{\check{\psi}}\gamma^{\mu}\partial_{\mu}\check{\psi}%
+\check{S}_{\mathrm{jA}}\,. \label{Spsi}%
\end{equation}
The $U(1)_{\star}$ gauge transformations read
\begin{align}
&  \check{A}_{\mu}\rightarrow\check{A}_{\mu}^{\prime}=U_{\check{\lambda}}%
\star\check{A}_{\mu}\star U_{\check{\lambda}}^{-1}+ig^{-1}\left(
\partial_{\mu}U_{\check{\lambda}}\right)  \star U_{\check{\lambda}}%
^{-1}\,,\nonumber\\
&  \check{F}_{\mu\nu}\rightarrow\check{F}_{\mu\nu}^{\prime}=U_{\check{\lambda
}}\star\check{F}_{\mu\nu}\star U_{\check{\lambda}}^{-1}\,,\nonumber\\
&  U_{\check{\lambda}}=e_{\star}^{i\check{\lambda}}=1+i\check{\lambda}%
-\frac{1}{2}\check{\lambda}\star\check{\lambda}+O\left(  \check{\lambda}%
^{3}\right)  \, \label{sw2}%
\end{align}
with a local parameter $\check{\lambda}(x)$. The action $\check{S}%
_{\mathrm{A}}$ is invariant under these gauge transformations (see e.g.,
\cite{SeiWit99,BicGriPopSchWul01,CarHarKosLanOka01}). The external current
$\ \check{j}^{\mu}\left(  x\right)  $ transforms covariantly,
\begin{equation}
\check{j}^{\mu\prime}=U_{\check{\lambda}}\star\check{j}^{\mu}\star
U_{\check{\lambda}}^{-1}\,. \label{sw2.2}%
\end{equation}
so that the equations of motion, $\delta\check{S}/\delta\check{A}_{\mu}=0$,%
\begin{equation}
\check{D}_{\nu}\check{F}^{\nu\mu}\left(  x\right)  =\frac{4\pi}{c}\check
{j}^{\mu}\left(  x\right)  \,, \label{sw2.3}%
\end{equation}
are gauge covariant under the transformations (\ref{sw2}), (\ref{sw2.2}). The
covariant derivative is defined as
\begin{equation}
\check{D}_{\mu}\Phi:=\partial_{\mu}\Phi+ig\left[  \check{A}_{\mu}%
\overset{\star}{,}\Phi\right]  \,. \label{covD}%
\end{equation}
The same transformation rule (\ref{sw2.2}) for the current follows from
(\ref{Spsi}) assuming standard transformation properties $\check{\psi
}\rightarrow\check{\psi}^{\prime}=U_{\check{\lambda}}\star\check{\psi}$ for
the fermions.

The compatibility condition for the equations of motion (\ref{sw2.3}) yields a
covariant conservation law for the current,
\begin{equation}
\check D_{\mu} \check D_{\nu}\check F^{\nu\mu}=0\Longrightarrow\check D_{\mu
}\check j^{\mu}=0\,. \label{2.2}%
\end{equation}

Let us now turn to the actions. The sum $\check{S}_{\mathrm{A}}+\check
{S}_{\check{\psi}}$ is $U(1)_{\star}$ invariant. Let us check what happens to
(\ref{SAjA}). By performing an infinitesimal gauge transformation (\ref{sw2}),
(\ref{sw2.2}) in the action for the currents we obtain,
\begin{equation}
\delta\check{S}_{\mathrm{jA}}=-\frac{1}{gc^{2}}\int dx\left\{  \left(
\partial_{\mu}\check{j}^{\mu}\right)  \star\check{\lambda}\right\}  \,,
\label{4}%
\end{equation}
which means that noncommutative Maxwell theory in presence of currents has
$U\left(  1\right)  _{\star}$ symmetry, if the noncommutative currents
$j^{\mu}$ are conserved $\partial_{\mu}j^{\mu}=0$. However, this disagrees
with the fact that the currents are covariantly conserved, which follows as an
identity from the equations of motion (\ref{2.2}). Hence, the total action
$\check{S}$ (\ref{SAjA}) is not gauge-invariant. In fact, this same feature is
already known for the Yang-Mills theory coupled with external currents. There
is no consistent way to introduce currents in nonabelian gauge theories
already at the classical level \cite{SikWei78} (see also \cite{Rammond})
without violating the gauge invariance, although a gauge-\textit{covariant}
theory can be formulated \cite{CabShab} both at classical and quantum levels.
Therefore, it is not unexpected that the same problem is encountered here due
to a close analogy between the $U\left(  1\right)  _{\star}$ noncommutative
gauge theory and nonabelian gauge theories \cite{Szabo,DouNek01}.

%%%%%%%%

\subsection{The Seiberg-Witten map}

\label{sec-SW} Despite the presence of all desired covariance properties of
the $U(1)_{\star}$ Maxwell equations, this is not what one needs to analyze
phenomenological predictions of the NC theory. One likes to deal with gauge
\emph{invariant} field strength and the currents rather than with gauge
\emph{covariant} ones. In other words, one one has to introduce ordinary
commutative $U(1)$ fields in place of the $U(1)_{\star}$ ones.

This is done with the help of the Seiberg-Witten map \cite{SeiWit99}, which is
very well known for $\check{A}_{\mu}$ and $\check{F}_{\mu\nu}$. At the first
order in $\theta$ it reads
\begin{align}
&  \check{A}_{\mu} =A_{\mu} +\frac{g} {2}\theta^{\alpha\beta}A_{\alpha}
\left[  \partial_{\beta}A_{\mu} +f_{\beta\mu} \right]  \,,\ \ f_{\mu\nu
}=\partial_{\mu}A_{\nu}-\partial_{\nu}A_{\mu}\,,\nonumber\\
&  \check{F}_{\mu\nu} =f_{\mu\nu} -g\theta^{\alpha\beta}\left[  f_{\alpha\mu}
f_{\beta\nu} -A_{\alpha} \partial_{\beta}f_{\mu\nu} \right]  \,.\label{sw4}\\
&  \check{\lambda}=\lambda-\frac g2 \theta^{\alpha\beta}\partial_{\alpha
}\lambda\cdot A_{\beta}\,.\nonumber
\end{align}
To find the SW map for currents $\check{j}^{\mu}$ one has to demand that under
the SW map the $U(1)_{*}$ gauge transformation (\ref{sw2.2}) of the current
$\check{j}^{\mu}$ is induced by the $U(1)$ gauge transformations of $A_{\mu}$
and $j^{\mu}$ through a functional dependence of $\check{j}^{\mu}$ on the
latter fields. For infinitesimal transformations this condition reads
\begin{align}
&  \check{j}^{\mu}\left(  A,j\right)  +\check{\delta}_{\check{\lambda}}%
\check{j}^{\mu}\left(  A,j\right)  =\check{j}^{\mu}\left(  A+\delta_{\lambda
}A,j\right)  \, \label{sw4.1}%
\end{align}
where $\check{\delta}_{\check{\lambda}}$ and ${\delta}_{{\lambda}}$ denote the
corresponding gauge variations. Note, that $\delta_{\lambda}j=0$, as it has to
be in the commutative electrodynamics. By virtue of (\ref{sw2.2})
\begin{align}
&  \check{\delta}_{\check{\lambda}}\check{j}^{\mu}\left(  A,j\right)
=i\left[  \check{\lambda}\overset{\star}{,}\check{j}^{\mu}\right]  =
-\theta^{\alpha\beta}\left(  \partial_{\alpha}\lambda\right)  \left(
\partial_{\beta}j^{\mu}\right)  +O\left(  \theta^{2}\right)  \,,
\label{sw4.1a}%
\end{align}
which coincides with the gauge transformation of $g\theta^{\alpha\beta
}A_{\alpha} \partial_{\beta}j^{\mu} $ caused by $\delta_{\lambda}A_{\alpha
}=-g^{-1}\partial_{\alpha}\lambda$. So, finally, the additional SW map for the
current is
\begin{equation}
\check{j}^{\mu} =j^{\mu} +g\theta^{\alpha\beta}A_{\alpha} \partial_{\beta
}j^{\mu} \,. \label{sw4.2}%
\end{equation}
This result coincides with the one derived previously in Ref.\ \cite{Ban}. The
SW map is not unique. There is a three-parameter ambiguity in the solutions of
the SW equations which will be discussed in Sec.\ \ref{sec-amb} below.

As it was already mentioned above, the NC action for electromagnetic field
interacting with an external current is \emph{not} gauge invariant, while the
field equations are gauge covariant. Therefore, it makes sense to apply the SW
map to the equation of motion (\ref{sw2.3}) and to the compatibility condition
(\ref{2.2}). To the first order in $\theta^{\mu\nu}$ one immediately gets
\begin{align}
&  \partial_{\nu} f^{\nu\mu}-g\theta^{\alpha\beta} \left(  \partial_{\nu
}(f_{\alpha}^{\ \nu}f_{\beta}^{\ \mu}) -f_{\nu\alpha}\partial_{\beta}f^{\nu
\mu} - A_{\alpha}\partial_{\nu}\partial_{\beta}f^{\nu\mu}\right)  =\frac{4\pi
}c (j^{\mu}+g\theta^{\alpha\beta}A_{\alpha} \partial_{\beta}j^{\mu
}),\label{SWeq}\\
&  \partial_{\mu}j^{\mu}+g\theta^{\alpha\beta} \left(  f_{\mu\alpha}%
\partial_{\beta}j^{\mu}+A_{\alpha}\partial_{\beta}\partial_{\mu}j^{\mu
}\right)  =0\,. \label{SWin}%
\end{align}
One can directly check the compatibility of these two equations within the
same $\theta$-accuracy by acting with $\partial_{\mu}$ on (\ref{SWeq}) and
using (\ref{SWin}) to get an identity. Only the antisymmetricity of the
tensors $\theta^{\alpha\beta}$ and $f^{\mu\nu}$ is referred to in the process,
as well as the Bianki identity for the latter.

The modified Maxwell equations (\ref{SWeq}) are nonlinear with respect to the
field already when the external current is away, $j=0$. The nonlinearity is
restricted to the second power of the field, because we confined ourselves to
the first order of the non-commutativity parameter $\theta$ when deducing
them: expansion in powers of $\theta$ generates expansion in powers of the field.

Equations of motion (\ref{SWeq}), (\ref{SWin}) are $U(1)$-gauge covariant by
construction, even though they look non-covariant as containing the potentials
together with the gauge-invariant field intensities and the current. It is
important that the potentials are involved with the small factor of $\theta$.
For this reason the potential-containing terms can be omitted from the set of
equations if taken on its solutions, determined within the accuracy accepted.
To prove this statement, note first that if the factor $\partial_{\mu}j^{\mu}%
$, which is of the order of $\theta$ according to equation (\ref{SWin}), is
substituted into its potential-containing term $g\theta^{\alpha\beta}%
A_{\alpha}\partial_{\beta}\partial_{\mu}j^{\mu},$ the latter becomes of the
order of $\theta^{2}$ and is, hence, to be disregarded in equation
(\ref{SWin}). Analogously, the difference of the two potential-containing
terms in the left- and right-hand sides of equation (\ref{SWeq}) $A_{\alpha
}~\theta^{\alpha\beta}\partial_{\beta}(\partial_{\nu}f^{\nu\mu}-\frac{4\pi}%
{c}j^{\mu})$ is also $\sim\theta^{2}$, because the factor $(\partial_{\nu
}f^{\nu\mu}-\frac{4\pi}{c}j^{\mu})$ in it is of the order of $\theta$
according to equation (\ref{SWeq}). Hence, we are left with the explicitly
gauge-invariant set of equations
\begin{align}
&  \partial_{\nu}f^{\nu\mu}-g\theta^{\alpha\beta}\left(  \partial_{\nu
}(f_{\alpha}^{\ \nu}f_{\beta}^{\ \mu})-f_{\nu\alpha}\partial_{\beta}f^{\nu\mu
}\right)  =\frac{4\pi}{c}j^{\mu},\label{SWeq0}\\
&  \partial_{\mu}j^{\mu}+g\theta^{\alpha\beta}f_{\mu\alpha}\partial_{\beta
}j^{\mu}=0\,. \label{SWin0}%
\end{align}

To treat the nonlinear equations (\ref{SWeq}), (\ref{SWin}) we shall need to
give them a recurrence form by expanding $A_{\mu}$ and $j^{\mu}$ in them in
the $\theta$-series :
\begin{align}
&  A_{\mu}=A_{\mu}^{\left(  0\right)  }+A_{\mu}^{\left(  1\right)  }\left(
\theta\right)  +O\left(  \theta^{2}\right)  \,,\label{sw8}\\
&  j^{\mu}=j^{(0)\mu}+j^{(1)\mu}+O\left(  \theta^{2}\right)  \,, \label{exj}%
\end{align}
where $A^{(0)}$, $j^{(0)}$ satisfy commutative Maxwell and current
conservation equations
\begin{equation}
\partial_{\nu}f^{\left(  0\right)  \nu\mu}=\frac{4\pi}{c}j^{(0)\mu}\,,
\qquad\partial_{\mu}j^{(0)\mu}=0\,, \label{Max0}%
\end{equation}
and $A^{(j)}$ and $j^{(j)}$ are corrections of the $j$th order in $\theta$. By
using (\ref{Max0}), we obtain to the first order in $\theta$
\begin{align}
&  \partial_{\nu} f^{(1)\nu\mu}-g\theta^{\alpha\beta} \left(  \partial_{\nu
}(f_{\alpha}^{(0)\nu}f_{\beta}^{(0)\mu}) -f_{\nu\alpha}^{(0)}\partial_{\beta
}f^{(0)\nu\mu} \right)  =\frac{4\pi}c j^{(1)\mu}\,,\label{SWeq1}\\
&  \partial_{\mu}j^{(1)\mu}+g\theta^{\alpha\beta} f_{\mu\alpha}^{(0)}%
\partial_{\beta}j^{(0)\mu}=0\,. \label{SWin1}%
\end{align}

Solutions of (\ref{SWeq1}) and (\ref{SWin1}) are not unique. One can add a
current $\tilde j^{\mu},$ which satisfies $\partial_{\mu}\tilde j^{\mu}=0$, to
$j^{(1)\mu}$. This is the same as equation (\ref{Max0}) for $j^{(0)\mu}$. This
ambiguity is, therefore, not a physical one and $\tilde j^{\mu}$ can be
absorbed in $j^{(0)\mu}$. In the examples considered in the next section,
where the source $j_{\mu}^{(0)}$ is static and spherically symmetric, one can
even take $j^{(1)\mu}=0$. Similarly, the ambiguity for $f^{(1)\mu\nu}$ can be
removed by imposing the fall-off conditions at infinity on this field or
fixing an external field.

Just for the sake of completeness, let us check what happens if one applies
the SW map at the level of the action (\ref{SAjA}). One easily obtains to the
first order in the NC parameter
\begin{align}
S_{\mathrm{SW}}  &  =-\frac{1}{16\pi c}\int dx\left\{  \left(  1+\frac{g}%
{2}\theta^{\alpha\beta}f_{\alpha\beta}\right)  f_{\mu\nu}f^{\mu\nu}%
-2g\theta^{\alpha\beta}f^{\mu\nu}f_{\alpha\mu}f_{\beta\nu}\right\} \nonumber\\
&  -\frac{1}{c^{2}}\int dx\left\{  j^{\mu}A_{\mu}+\frac{g}{2}\theta
^{\alpha\beta}j^{\mu}A_{\alpha}\left(  \partial_{\beta}A_{\mu}+f_{\beta\mu
}\right)  +g\theta^{\alpha\beta}A_{\mu}A_{\alpha}\left(  \partial_{\beta
}j^{\mu}\right)  \right\}  \,. \label{sw5.1}%
\end{align}
Equating to zero the variation of this action with respect to $A_{\mu}$ yields
the following equation
\begin{align}
\partial_{\nu}\left[  f^{\nu\mu}\left(  1+\frac{g}{2}\theta^{\alpha\beta
}f_{\alpha\beta}\right)  \right]   &  =\frac{4\pi}{c}j^{\mu}\left(  1+\frac
{g}{2}\theta^{\alpha\beta}f_{\alpha\beta}\right)  +g\theta^{\alpha\beta
}\left[  \partial_{\nu}\left(  f_{\alpha}^{\ \nu}f_{\beta}^{\ \mu}\right)
+\partial_{\alpha}\left(  f^{\mu\nu}f_{\beta\nu}\right)  \right] \nonumber\\
&  +g\theta^{\beta\mu}\left\{  \partial_{\alpha}\left(  f^{\alpha\nu}%
f_{\beta\nu}\right)  -\frac{1}{4}\partial_{\beta}\left(  f^{\alpha\nu
}f_{\alpha\nu}\right)  \right. \nonumber\\
&  -\left.  \frac{4\pi}{c}\left[  f_{\beta\alpha}j^{\alpha}+A_{\alpha}\left(
\partial_{\beta}j^{\alpha}\right)  - \frac{A_{\beta}}{2}\left(  \partial
_{\alpha}j^{\alpha}\right)  \right]  \right\}  \,. \label{sw5.3}%
\end{align}
When the current is away, it can be directly checked that these equations
coincide with (\ref{SWeq}) on solutions of the latter. On the contrary, when
$j^{\mu}\neq0$, equations (\ref{sw5.3}) exhibit troubles. The gauge invariance
of (\ref{sw5.3}) cannot be restored even by doing the expansions (\ref{sw8})
and (\ref{exj}). In what follows, we discard the action approach, and use
exclusively the NC Maxwell and current conservation equations (\ref{SWeq}) and
(\ref{SWin}) (together with their expanded versions (\ref{SWeq1}) and
(\ref{SWin1})).

Some clarifying remarks on the non-equivalence of the SW maps in the action
and in equations of motion are in order. By definition, for an $U(1)_{\star}$
theory, $\check{S}(\check{A},\check{j})=S_{\mathrm{SW}}(A,j)$. Therefore, the
equation of motion (\ref{sw5.3}) can be rewritten as
\begin{equation}
0=\frac{\delta S_{\mathrm{SW}}}{\delta A_{\mu}(x)}=\int dy\left[  \frac
{\delta\check{S}}{\delta\check{A}_{\nu}(y)}\frac{\delta\check{A}_{\nu}%
(y)}{\delta A_{\mu}(x)}+\frac{\delta\check{S}}{\delta\check{j}^{\nu}(y)}%
\frac{\delta\check{j}^{\nu}(y)}{\delta A_{\mu}(x)}\right]  \,. \label{SWS}%
\end{equation}
The first term in the brackets above vanishes on the equations of motion
(\ref{sw2.3}), while the second one does not (since there is no equation of
motion produced by variations of currents). Hence, the equations of motion
obtained by varying $S_{\mathrm{SW}}$ are not equivalent to that obtained from
the original action $\check{S}$. On the opposite, if the SW map is applied to
the equations of motion of the $U(1)_{\star}$ gauge theory, such an
equivalence is, of course, preserved (after truncation to the first order in
$\theta$). The reason for non-equivalence of the two procedures is in the
non-dynamical nature of the external current, which does not generate any
equations of motion, but participates in the SW map.

%%%%%%%%%%%

\section{Solutions for the potential produced by a static charge in the
presence of a magnetic background}

\label{sec-mag}

\subsection{Regular solutions}

\label{subsection3.1}

In this section we are studying the linear in $\theta$ NC corrections to the
4-vector potential of a static spherically symmetric charge. The corrections
to be found are both magnetostatic and electrostatic (the latter will occur
only when an external magnetic field is present).

We impose the stationarity conditions
\begin{equation}
\partial_{0}A^{\left(  1\right)  \mu}\left(  x\right)  =0\,, \label{sw10}%
\end{equation}
on these corrections, bearing in mind that the unperturbed solutions, i.e.
those to the equations
\begin{align}
&  \partial_{\nu}f^{\left(  0\right)  \nu\mu}=\frac{4\pi}{c}j^{\left(
0\right)  \mu}\,,\ \ \partial_{\mu}j^{\left(  0\right)  \mu}=0\,,\nonumber\\
&  f_{\mu\nu}^{\left(  0\right)  }=\partial_{\mu}A_{\nu}^{\left(  0\right)
}-\partial_{\nu}A_{\mu}^{\left(  0\right)  }\,, \label{sw9}%
\end{align}
are also subjected to the stationarity conditions. More precisely, we take the
total external charge $Ze$ distributed with a constant density throughout a
spherical region of the space with the radius $a$ . The current density
$j^{\left(  0\right)  \mu}$ is defined in the inner part of the sphere $r<a$,
called region $\mathrm{I}$, and in its outer part $r>a$, called region
$\mathrm{II}$, as follows
\begin{align}
&  j^{\mu}=\left(  c\rho,0\right)  \,,\ \ \rho\left(  \mathbf{x}\right)
=\left\{
\begin{array}
[c]{l}%
\frac{3}{4\pi}\frac{Ze}{a^{3}}\,,\ \ r\in\mathrm{I}\\
0\,,\ \ r\in\mathrm{II}%
\end{array}
\,,\ \ r=\left\vert \mathbf{x}\right\vert \ \right.  \label{10.1}%
\end{align}
In what follows we shall equip designations of potentials, relating to these
regions, with the indices $\mathrm{I}$ or $\mathrm{II}$. The homogeneous --
inside the sphere -- charge distribution is taken for simplicity. Extensions
to arbitrary spherical symmetric distributions, continuous ones included, may
be also considered when necessary. The charge density (\ref{10.1}) tends to
the Dirac delta-function in the point-charge limit: $\rho(\mathbf{x}) =
Ze~\delta^{3}(\mathbf{x})$, as $a\to0$. However, owing to the coordinate
noncommutativity, different coordinate components cannot be simultaneously
given definite values, hence no spherical \textit{physical object} should be
taken with its radius smaller than the elementary length. For this reason we
will restrict our consideration to the values $a>\sqrt{\theta}$. On the other
hand, after the SW map is accomplished, we deal with a commutative plain and
must take care of providing consistency to the resulting theory defined
everywhere on it, both for $r<a$ and $r>a$. Therefore, when considering the
coordinate values down to the origin $r=0$, we must seriously treat possible
singularities in this point and their consequences. This will lead us to
imposing regularity boundary conditions in the origin. This is a must at least
as long as the $\theta$-expansion is relied on. The point is that higher
orders of $\theta$ are accompanied by higher orders of the electromagnetic
potential and its derivatives. Therefore, if a singularity is admitted in the
lowest order, it would strengthen with every next order of the $\theta
$-expansion, hence the latter would not exist.

Note, that in Ref.\ \cite{SmSp} it was even suggested that the smearing of
charges replaces the use of noncommutative products in equations of motion. We
do not share this point of view.

Eq. (\ref{10.1}), certainly, satisfies the current-conservation equation in
(\ref{sw9}). Equation (\ref{sw9}) is satisfied by the following
electromagnetic potential $A^{\left(  0\right)  \mu}$,
\begin{align}
&  A^{\left(  0\right)  \mu}=\left(  A^{\left(  0\right)  0},A^{\left(
0\right)  i}\right)  \,,\nonumber\\
&  A^{\left(  0\right)  0}\left(  r\right)  =\left\{
\begin{array}
[c]{l}%
-\frac{Ze}{2a^{3}}r^{2}+\frac{3}{2}\frac{Ze}{a}\,,\ \ r\in\mathrm{I}\\
\frac{Ze}{r}\,,\ \ r\in\mathrm{II}%
\end{array}
\right.  \,,\quad A^{(0)i}=-\frac1{2}f^{(0)}_{ik}x^{k}\,,\quad f_{ij}^{\left(
0\right)  }=\mathrm{const}\,, \label{13}%
\end{align}
where we have included a solution of the homogeneous counterpart of equation
(\ref{sw9}) $A^{(0)i}=-\frac1{2}f^{(0)}_{ik}x^{k}$ corresponding to a constant
external magnetic field $B_{i}=\frac1{2}\varepsilon_{ijk}f^{(0)}_{jk}$. The
case $A^{(0)i}=0$ has been considered previously \cite{AdoGitShaVas11}.

The zeroth component of equation (\ref{13}) satisfies the boundary
\begin{equation}
A_{\mathrm{I}}^{\left(  0\right)  0}\left(  0\right)  \neq\infty,\qquad\left.
A_{\mathrm{II}}^{\left(  0\right)  0}\left(  r\right)  \right\vert
_{r\rightarrow+\infty}=0\,, \label{bc}%
\end{equation}
and smoothness $\left.  A_{\mathrm{I}}^{\left(  0\right)  0}\left(  r\right)
\right\vert _{r=a}=\left.  A_{\mathrm{II}}^{\left(  0\right)  0}\left(
r\right)  \right\vert _{r=a}$, $\left.  \partial_{r}A_{\mathrm{I}}^{\left(
0\right)  0}\left(  r\right)  \right\vert _{r=a}=\left.  \partial
_{r}A_{\mathrm{II}}^{\left(  0\right)  0}\left(  r\right)  \right\vert _{r=a}$
conditions. The boundary conditions (\ref{bc}) completely determine the
solution (\ref{13}) for $A^{(0)0}$ of the Laplace equation. The second
condition in (\ref{bc}) excludes the linear function $E^{i}x^{i}$
corresponding to a homogeneous electric field of arbitrary strength and of
arbitrary direction as a possible solution for $f^{\left(  0\right)  0i}$.

We restrict ourselves to a space-space noncommutativity $\left(  \theta^{0\mu
}=0\right)  $. From the spherical symmetry of the current $j^{\left(
0\right)  \mu}$ (\ref{10.1}) and of the solution $A^{(0)0}$ (\ref{13}) it
follows that equation (\ref{SWin1}) is satisfied by $j^{\left(  1\right)  \mu
}=0$, no correction to the current is required. This implies that the current
remains dynamically intact, $j^{\mu}=j^{\left(  0\right)  \mu}$, so we may
refer to it as a fixed external current, as this is customary in an
$U(1)$-theory. The Maxwell equation in SW approach (\ref{SWeq1}) reads,%
\begin{equation}
\partial_{i}f^{\left(  1\right)  i\mu}+g\theta^{ij}\left[  \left(
\partial_{i}A^{\left(  0\right)  0}\right)  \left(  \partial_{j}\partial^{\mu
}A^{\left(  0\right)  0}\right)  +2f_{ik}^{\left(  0\right)  }\left(
\partial_{k}\partial_{j}A^{\left(  0\right)  \mu}\right)  \right]
=0\,.\label{13.1}%
\end{equation}
Taking into account the stationarity condition, for the zeroth component
$\left(  \mu=0\right)  $, eq. (\ref{13.1}) reduces to
\begin{equation}
\boldsymbol{\nabla}^{2}A^{\left(  1\right)  0}+2gB^{i}\theta^{j}\left(
\delta_{ij}\boldsymbol{\nabla}^{2}A^{\left(  0\right)  0}-\partial_{i}%
\partial_{j}A^{\left(  0\right)  0}\right)  =0\,,\label{16.1}%
\end{equation}
where we have introduced the vector $\boldsymbol{\theta}$ with components
$\theta^{i}=\frac{1}{2}\varepsilon_{ijk}\theta^{jk}$. The spacial part of eq.
(\ref{13.1}) reads
\begin{align}
\partial_{i}f^{\left(  1\right)  ik}+g\theta^{ij}\left(  \partial
_{i}A^{\left(  0\right)  0}\right)  \left(  \partial^{k}\partial_{j}A^{\left(
0\right)  0}\right)   &  =0\nonumber\\
&  \mathrm{or}\hspace{10cm}\nonumber\\
\mathbf{\nabla}^{2}A^{\left(  1\right)  k}-\partial_{i}\partial_{k}A^{\left(
1\right)  i}+g\theta^{ij}\left(  \partial_{i}A^{\left(  0\right)  0}\right)
\left(  \partial^{k}\partial_{j}A^{\left(  0\right)  0}\right)   &
=0\,,\label{11.2}%
\end{align}
because the second space derivative of the external constant magnetic field
potential $A^{(0)k}$ is zero.

Once the starting equations (\ref{SWeq1}), (\ref{SWin1}) contain only field
strengthes and not potentials we may impose, for instance, the Coulomb gauge
condition $\partial_{k}A^{k}=0$ both on $A^{(0)k}$ and on $A^{(1)k}$. It is
worthwhile making explicitly sure that the resulting equations are
noncontradictory. To this end, let us act with the differential operator
$\partial_{k}$ on (\ref{11.2}) and see that we do not come to a contradiction
with the equality $\partial_{k}A^{(1)k}$ = 0. This implies the requirement
$\theta^{ij}\partial_{k}[(\partial_{i}A^{(0)0})(\partial_{k}\partial
_{j}A^{(0)0})]=0$. This expression is equal to $\theta^{ij}[(\partial
_{k}\partial_{i}A^{(0)0})(\partial_{k}\partial_{j}A^{(0)0})+\left(
\partial_{i}A^{(0)0}\right)  (\partial_{k}^{2}\partial_{j}A^{(0)0})]$. The
first term here disappears due to the antisymmetricity of $\theta^{ij}$. The
second one is zero thanks to the spherical symmetry of $A_{(0)0}$, already
exploited above, because the latter implies that this scalar function contains
only one 3-vector, which is the radius-vector $\mathbf{x}$. Then the product
of the two factors in the brackets is proportional to the product of different
components of this same vector $x_{i}x_{j}$. This tensor disappears when
multiplied by the antisymmetric tensor $\theta^{ij}$. In the Coulomb gauge we
obtain finally for spacial components $\left(  \mu=k=1,2,3\right)  $
\begin{equation}
\boldsymbol{\nabla}^{2}A^{\left(  1\right)  k}+g\theta^{ij}\left(
\partial_{i}A^{\left(  0\right)  0}\right)  \left(  \partial_{k}\partial
_{j}A^{\left(  0\right)  0}\right)  =0\,. \label{12.2}%
\end{equation}
One can observe that unlike eq. (\ref{16.1}), the latter equation (\ref{12.2})
does not contain a contribution from the external magnetic field. These facts
show that the introduction of a constant and homogeneous magnetic background
will not modify the magnetic field produced by a static charge, but instead,
provides a correction to its electric field, which otherwise ($\mathbf{B=0}$)
does not gain first-order $\theta$-correction, since equation (\ref{16.1}) for
it becomes source-free Laplace equation with the regular boundary conditions
(\ref{bc}).

We see that equations for $A^{(1)0}$ and $A^{(1)i}$ decouple and can be
analyzed separately. We start with the equation for $A^{(1)i}$. In region
$\mathrm{I}$, equation (\ref{12.2}) reads,%
\begin{equation}
\mathbf{\nabla}^{2}A_{\mathrm{I}}^{\left(  1\right)  k}\left(  \mathbf{x}%
\right)  =-g\left(  \frac{Ze}{a^{3}}\right)  ^{2}\theta^{ik}x^{i}\,,
\label{es1}%
\end{equation}
and for region $\mathrm{II}$ we have
\begin{equation}
\mathbf{\nabla}^{2}A_{\mathrm{II}}^{\left(  1\right)  k}\left(  \mathbf{x}%
\right)  =-g\left(  \frac{Ze}{r^{3}}\right)  ^{2}\theta^{ik}x^{i}\,.
\label{es2}%
\end{equation}

The general solutions are%
\begin{align}
A_{\mathrm{I}}^{\left(  1\right)  k}\left(  \mathbf{x}\right)   &  =-\frac
{g}{10}\left(  \frac{Ze}{a^{3}}\right)  ^{2}r^{2}\theta^{ik}x^{i}%
+a_{\mathrm{I}}^{\left(  1\right)  k}\left(  r,\vartheta,\varphi\right)
\,,\label{es3.1}\\
A_{\mathrm{II}}^{\left(  1\right)  k}\left(  \mathbf{x}\right)   &  =-\frac
{g}{4}\left(  \frac{Ze}{r^{2}}\right)  ^{2}\theta^{ik}x^{i}+a_{\mathrm{II}%
}^{\left(  1\right)  k}\left(  r,\vartheta,\varphi\right)  \,, \label{es3}%
\end{align}
where $\vartheta$ and $\varphi$ are the azimuthal and polar angles of the
coordinate radius-vector $\mathbf{x}$, respectively. The functions
$a_{\lambda}^{\left(  1\right)  k}$ $\left(  \lambda=\mathrm{I},\mathrm{II}%
\right)  $\ are solutions of the homogeneous Laplace equation
$\boldsymbol{\nabla}^{2}a_{\lambda}^{\left(  1\right)  k}=0$,%
\[
a_{\lambda}^{\left(  1\right)  k}\left(  r,\vartheta,\varphi\right)
=\sum_{l=0}^{\infty}\sum_{m=-l}^{+l}\left[  \alpha_{\left(  \lambda\right)
l,m}^{k}r^{l}+\beta_{\left(  \lambda\right)  l,m}^{k}r^{-\left(  l+1\right)
}\right]  Y_{l,m}\left(  \vartheta,\varphi\right)  \,,
\]
where the functions $Y_{l,m}\left(  \vartheta,\varphi\right)  $ are spherical
Harmonics \cite{Jackson}. The constants $\alpha_{\left(  \lambda\right)
l,m}^{k}$ and $\beta_{\left(  \lambda\right)  l.m}^{k}$ are fixed by the same
type of boundary and smoothness conditions as before,
\begin{align}
&  \left.  \mathbf{A}_{\mathrm{I}}^{\left(  1\right)  }\left(  \mathbf{x}%
\right)  \right\vert _{r=a}=\left.  \mathbf{A}_{\mathrm{II}}^{\left(
1\right)  }\left(  \mathbf{x}\right)  \right\vert _{r=a}\,,\nonumber\\
&  \left.  \frac{\partial}{\partial r}\mathbf{A}_{\mathrm{I}}^{\left(
1\right)  }\left(  \mathbf{x}\right)  \right\vert _{r=a}=\left.
\frac{\partial}{\partial r}\mathbf{A}_{\mathrm{II}}^{\left(  1\right)
}\left(  \mathbf{x}\right)  \right\vert _{r=a}\,,\nonumber\\
&  \left.  \mathbf{A}_{\mathrm{I}}^{\left(  1\right)  }\left(  \mathbf{x}%
\right)  \right\vert _{r\rightarrow0}\neq\infty,\qquad\left.  \mathbf{A}%
_{\mathrm{II}}^{\left(  1\right)  }\left(  \mathbf{x}\right)  \right\vert
_{r\rightarrow\infty}=0\,. \label{es4}%
\end{align}
The right-hand sides in eqs. (\ref{es1}) and (\ref{es2}) can be expressed in
terms of spherical harmonics with $l=1$ using the relations
%\cite{Jackson},%
\begin{equation}
Y_{1,\pm1}=\mp\frac{1}{r}\sqrt{\frac{3}{8\pi}}\left(  x^{1}\pm ix^{2}\right)
\,,\ \ Y_{1,0}=\frac{1}{r}\sqrt{\frac{3}{4\pi}}x^{3}\,. \label{rel}%
\end{equation}
Then the coefficients $\alpha_{\left(  \lambda\right)  l,m}^{k}$ and
$\beta_{\left(  \lambda\right)  l,m}^{k}$ are found to be ($\alpha_{\left(
\mathrm{I}\right)  l,m}^{k}=\alpha_{l,m}^{k},$ $\beta_{\left(  \mathrm{II}%
\right)  l,m}^{k}=\beta_{l,m}^{k}$)
\begin{align}
&  \alpha_{\left(  \mathrm{II}\right)  l,m}^{k}=\beta_{\left(  \mathrm{I}%
\right)  l,m}^{k}=0\,,\nonumber\\
&  \alpha_{1,\pm1}^{k}=\frac{1}{4}g\left(  \frac{Ze}{a^{2}}\right)  ^{2}%
\sqrt{\frac{2\pi}{3}}\left[  \mp\theta^{1k}+i\theta^{2k}\right]
\,,\nonumber\\
&  \beta_{1,\pm1}^{k}=\frac{2}{5}g\frac{\left(  Ze\right)  ^{2}}{a}\sqrt
{\frac{2\pi}{3}}\left[  \mp\theta^{1k}+i\theta^{2k}\right]  \,,\nonumber\\
&  \alpha_{1,0}^{k}=\frac{1}{4}g\left(  \frac{Ze}{a^{2}}\right)  ^{2}%
\sqrt{\frac{4\pi}{3}}\theta^{3k}\,,\nonumber\\
&  \beta_{1,0}^{k}=\frac{2}{5}g\frac{\left(  Ze\right)  ^{2}}{a}\sqrt
{\frac{4\pi}{3}}\theta^{3k}\,, \label{8}%
\end{align}
while all the rest of the coefficients $\alpha_{l,m}^{k}$ and $\beta_{l.m}%
^{k}$ with $l\neq1$ are identically zero. Finally we have
\begin{align}
&  A_{\mathrm{I}}^{\left(  1\right)  k}\left(  \mathbf{x}\right)  =-\frac
{g}{4}\left(  \frac{Ze}{a^{2}}\right)  ^{2}\left(  \frac{2}{5}\frac{r^{2}%
}{a^{2}}-1\right)  \theta^{ik}x^{i}\,,\quad r<a\,,\nonumber\\
&  A_{\mathrm{II}}^{\left(  1\right)  k}\left(  \mathbf{x}\right)  =\frac
{g}{4}\left(  \frac{Ze}{r^{2}}\right)  ^{2}\left(  \frac{8}{5}\frac{r}%
{a}-1\right)  \theta^{ik}x^{i}\,,\quad r>a\,. \label{es5}%
\end{align}
As a matter of fact, this solution disappears in the origin $\mathbf{A}%
_{\mathrm{I}}^{\left(  1\right)  }\left(  0\right)  =0$, although such a
boundary condition was not imposed. There is no finite limit of (\ref{es5}) if
$a\rightarrow0$.

Now we proceed to evaluating the solutions of (\ref{16.1}). For region
$\mathrm{I}$ we obtain%
\begin{equation}
\boldsymbol{\nabla}^{2}A_{\mathrm{I}}^{\left(  1\right)  0}\left(
\mathbf{x}\right)  =4g\left(  \frac{Ze}{a^{3}}\right)  \left(  \mathbf{B}%
\cdot\boldsymbol{\theta}\right)  \,, \label{sw21}%
\end{equation}
and for region $\mathrm{II}$ we have%
\begin{equation}
\boldsymbol{\nabla}^{2}A_{\mathrm{II}}^{\left(  1\right)  0}\left(
\mathbf{x}\right)  =2g\left(  \frac{Ze}{r^{5}}\right)  \left[  3\left(
\mathbf{x}\cdot\mathbf{B}\right)  \left(  \mathbf{x}\cdot\boldsymbol{\theta
}\right)  -r^{2}\left(  \mathbf{B}\cdot\boldsymbol{\theta}\right)  \right]
\,, \label{sw22}%
\end{equation}
we remind that the vector $\boldsymbol{\theta}$ has components $\theta
^{i}=\frac{1}{2}\varepsilon_{ijk} \theta^{jk}$.

The general solutions can be expressed as%
\begin{align}
A_{\mathrm{I}}^{\left(  1\right)  0}\left(  \mathbf{x}\right)   &  =\frac
{2}{3}g\left(  \frac{Ze}{a^{3}}\right)  r^{2}\left(  \mathbf{B}\cdot
\boldsymbol{\theta}\right)  +a_{\mathrm{I}}^{\left(  1\right)  0}\left(
r,\vartheta,\varphi\right)  \,,\label{sw23.1}\\
A_{\mathrm{II}}^{\left(  1\right)  0}\left(  \mathbf{x}\right)   &  =-g\left(
Ze\right)  \frac{\left(  \mathbf{x}\cdot\mathbf{B}\right)  \left(
\mathbf{x}\cdot\boldsymbol{\theta}\right)  }{r^{3}}+a_{\mathrm{II}}^{\left(
1\right)  0}\left(  r,\vartheta,\varphi\right)  \,, \label{sw23}%
\end{align}
where $a_{\mathrm{I}}^{\left(  1\right)  0},a_{\mathrm{II}}^{\left(  1\right)
0}$ are homogeneous solutions, which are fixed through the boundary conditions
(\ref{es4}) yielding%
\begin{align}
&  A_{\mathrm{I}}^{\left(  1\right)  0}\left(  \mathbf{x}\right)  =2g\left(
\frac{Ze}{a}\right)  \left\{  \left(  \frac{2}{5}\frac{r^{2}}{a^{2}}-1\right)
\left(  \mathbf{B}\cdot\boldsymbol{\theta}\right)  -\frac{1}{5a^{2}}\left(
\mathbf{x}\cdot\mathbf{B}\right)  \left(  \mathbf{x}\cdot\boldsymbol{\theta
}\right)  \right\}  \,,\quad r<a\,,\nonumber\\
&  A_{\mathrm{II}}^{\left(  1\right)  0}\left(  \mathbf{x}\right)  =g\left(
\frac{Ze}{r}\right)  \left\{  \frac{1}{r^{2}}\left(  \frac{3}{5}\frac{a^{2}%
}{r^{2}}-1\right)  \left(  \mathbf{x}\cdot\mathbf{B}\right)  \left(
\mathbf{x}\cdot\boldsymbol{\theta}\right)  -\left(  \frac{1}{5}\frac{a^{2}%
}{r^{2}}+1\right)  \left(  \mathbf{B}\cdot\boldsymbol{\theta}\right)
\right\}  \,\quad r>a\,. \label{sw24}%
\end{align}

By using (\ref{sw4}) and (\ref{sw4.2}) together with the solutions obtained
above one can obtain the noncommutative fields $\check{A}^{\mu}$ and currents
$\check{j}^{\mu}$ at the same (linear) order in $\theta$. As expected,
noncommutative fields differ from their SW counterparts already at this order.

It might make sense to confront the result (\ref{sw24}) with the fact
\cite{GorMir} that within certain models the large magnetic field regime
mimics the noncommutivity: the (effective) action for some composite or gauge
fields includes their Moyal-like product, so these fields may be imagined as
defined on the coordinates, out of which the ones, orthogonal to the magnetic
field, do not mutually commute. To attribute the origin of noncommutativity
dealt with in the present context to this effect we have to identify the
elementary length with the Larmour radius, $\sqrt{\theta}=1/\sqrt{eB},$ taking
the magnetic field large and coinciding with $\boldsymbol{\theta}$ in
direction. Then, $B=\infty$ is the commutative limit, while the dimensionality
of space, $d=4$, is reduced by two -- the number of coordinates, orthogonal to
$\mathbf{B}$: the orthogonal subspace merely does not exist. On the contrary,
while $B$ is large, but still finite, the orthogonal subspace is
noncommutative, while the total space is \textquotedblleft
almost\textquotedblright\ $d-2=2$ dimensional. (The reservation
\textquotedblleft almost\textquotedblright\ means that in some domains of the
space, say, near a charge, where its field dominates over the external
magnetic field, the dimensionality of space is again \cite{ShaUs07} $d=4$).
Unfortunately, the result (\ref{sw24}) cannot cover this case, because the
condition $B\theta=1$ lies beyond the applicability of the expansion in powers
of $\theta$ and $B$, used in deriving eq. (\ref{sw24}).
%%%%%%%%%%%%%%

\subsection{Alternative magnetic solutions}

\label{subsection3.2} The solutions that we obtained above depend crucially on
the regularity condition imposed at $r=0$. Relaxing this condition, one can
obtain another solution for the vector-potential
\begin{align}
A_{\mathrm{I}}^{\left(  1\right)  k}\left(  \mathbf{x}\right)   &  =-\frac
{g}{4}\left(  \frac{Ze}{a^{2}}\right)  ^{2}\left(  \frac{2}{5}\frac{r^{2}%
}{a^{2}}+\frac{8}{5}\frac{a^{3}}{r^{3}}-1\right)  \theta^{ik}x^{i}%
\,,\nonumber\\
A_{\mathrm{II}}^{\left(  1\right)  k}\left(  \mathbf{x}\right)   &  =-\frac
{g}{4}\left(  \frac{Ze}{r^{2}}\right)  ^{2}\theta^{ik}x^{i}\, \label{15}%
\end{align}
that does not obey the finiteness boundary condition in the origin, but
decreases at large distance from the source faster than (\ref{es5}), in other
words it is short-range. Its outer part, $A_{\mathrm{II}}^{(1)k}(\mathbf{x),}$
which does not depend on the size $a$ of the charge, coincides with the
magnetic solution found in \cite{Stern} for the field produced by a point-like
static charge (the limit of $a=0$ in (\ref{10.1})); it is highly singular in
the origin $r=0$ in that case. Unlike eq. (\ref{es5}) this solution is not the
field of a magnetic dipole, since it decreases at large distances faster than that.

Solution (\ref{es5}) differs from (\ref{15}) by adding the function
\begin{equation}
2g\frac{(Ze)^{2}}{5a}\frac{\theta^{ik}x^{i}}{r^{3}}, \label{free}%
\end{equation}
which is a solution for the vector potential to the homogeneous Laplace
equation with the constant coefficient $2g\frac{(Ze)^{2}}{5a}$ chosen in such
a way as to cancel the singularity in the origin of (the inner part of) the
solution (\ref{15}). There are three more homogeneous solutions that do not
include any vectors or tensors, other than the ones inherent to the problem:
the radius-vector $\mathbf{x}$ and the noncommutativity tensor $\theta^{ik}$:
\begin{equation}
\frac{\theta^{k}}{r},\qquad\qquad\frac{x_{k}}{r^{3}},\qquad\qquad\theta
^{ki}x_{i} . \label{three}%
\end{equation}
The first one does not satisfy the Coulomb gauge condition and should not,
therefore, be included into consideration. The second one is a pure gauge, it
does not contain any field strength in it. Its appearance is due to the fact
that the Coulomb gauge fixes the gauge degree of freedom only up to a gauge
transformation caused by a function $\lambda(r)$ obeying the free Laplace
equation. So, with $\lambda=1/r$ the discussed solution is $-\partial
_{k}\lambda$. Thus, only the third solution $\theta^{ki}x_{i}$ remains yet to
be considered. It is linear in $\mathbf{x}$. Once the smoothness conditions in
(\ref{es4}) include the first and the second derivatives, the linear solution
can be matched only with itself on the internal boundary. The third
homogeneous solution cannot be associated with the source, a constant
coefficient to be put in front of it remaining arbitrary. We conclude that the
two solutions (\ref{15}) and (\ref{es5}) exhaust all magnetic solutions
produced by the static charge. There are, certainly, many more, homogeneous
solutions not associated with the charge. One of them is $\theta^{ki}x_{i}$.
It corresponds to a constant and homogeneous magnetic field of arbitrary
strength, but of fixed direction: it is directed along the noncommutativity
vector $\boldsymbol{\theta}$. This field should be absorbed into the (more
general) external magnetic field $\mathbf{B}$, already included in (\ref{13})
at the zero-order level. Note that the total magnetic energy of the static
charge $\simeq\int|\mathbf{rot A}|^{2}\mathrm{d}^{3}x$ is finite for
(\ref{es5}) and infinite for (\ref{15}) when $a$ is finite.

Which of the two solutions (\ref{15}) or (\ref{es5}) should be selected? Let
us first seek an answer to this question beyond the intrinsic context of the
NC theory, taking into consideration a possible future use of the solution.
According to \cite{Stern} the magnetic field carried by solutions for the
3-vector potential interact with the orbital momentum and the spin of the
electron in a NC hydrogen atom problem elaborated in \cite{ChaiSheikhTur},
wherein the noncommutative charge is taken for the nucleus and placed in the
origin $r=0$. This interaction energy, computed using our solution (\ref{es5})
is finite in the origin. On the contrary, it behaves as $r^{-3}$ for solution
(\ref{15}) (and even worse, as $r^{-4},$ if the outer part of (\ref{15}) is
continued to the origin to form the point charge solution of \cite{Stern}).
The contribution of this interaction energy causes the fall-down onto the
center and thus makes the problem inconsistent. Nevertheless, the similar
situation is not considered (although not quite righteously) as a real trouble
in quantum mechanics, because the finite size of the nucleus provides a
sufficient cut-off. So, purely pragmatically, we cannot completely justify the
exclusive necessity of selecting solution (\ref{es5}), but we shall come back
to this discussion later in this section after considering also solutions
alternative to the electric solution (\ref{sw24}).

\subsection{Alternative electrostatic solutions}

\label{subsection3.3} A clue observation that has helped to solve equation
(\ref{sw21}), (\ref{sw22}) and may be used to check its solution (\ref{sw24})
is that a linear combination $a(r)(\boldsymbol{B\cdot\theta}%
)+b(r)(\mathbf{x\cdot B})(\mathbf{x}\boldsymbol{\cdot\theta})$ reproduces
itself with different coefficients after being acted on by the Laplace
operator according to the formulae:%
\begin{align}
\boldsymbol{\nabla}^{2}a(r)(\boldsymbol{B\cdot\theta})=(\boldsymbol{B\cdot
\theta}) \left(  a^{\prime\prime}+ \frac{2a^{\prime}}{r}\right)  ,\nonumber\\
\boldsymbol{\nabla}^{2}b(r)(\mathbf{x\cdot B})(\mathbf{x}\boldsymbol{\cdot
\theta})=\left(  b^{\prime\prime}+\frac{6b^{\prime}}{r}\right)
(\mathbf{x\cdot B})(\mathbf{x}\boldsymbol{\cdot\theta})+
2b(r)(\boldsymbol{B\cdot\theta}), \label{clue}%
\end{align}
where the primes denote differentiations over $r$. Using eq. (\ref{clue}) and
the general solution \cite{Kamke}
\begin{equation}
y(x)=\left(  \frac\xi{x}\right)  ^{\gamma}\left(  \eta+ \int_{\xi}%
^{x}g(x^{\prime})\left(  \frac{x^{\prime}}{\gamma}\right)  ^{\gamma}%
\mathrm{d}x^{\prime}\right)
\end{equation}
to the linear first-order differential equation%
\begin{equation}
y^{\prime}+\frac{\gamma}{x}y=g(x),
\end{equation}
where $\xi$ and $\eta$ are arbitrary constants, and $\gamma$ is either 2 or 6
in our case, one can find all solutions to equations (\ref{sw21}),
(\ref{sw22}) of the form $a(r)(\boldsymbol{B\cdot\theta})+b(r)(\mathbf{x\cdot
B})(\mathbf{x}\boldsymbol{\cdot\theta}),$ extending, when necessary, beyond
the homogeneous charge distribution (\ref{10.1}). (The reason to confine
ourselves to this class of solutions is that, once the inhomogeneity in
(\ref{sw21}), (\ref{sw22}) is linear both in $\mathbf{B}$ and
$\boldsymbol{\theta}$, only solutions with the same property may be referred
to as produced by the source under consideration). The homogeneous solutions
of equations (\ref{sw21}), (\ref{sw22}) so found are
\begin{equation}
\mathrm{a)}\;\;(\boldsymbol{B\cdot\theta}),\quad\mathrm{b)}\;\; \frac
{(\boldsymbol{B\cdot\theta})}{r}, \quad\mathrm{c)}\;\; \frac
{(\boldsymbol{B\cdot\theta})}{r^{3}}-\frac{3(\mathbf{x\cdot B})(\mathbf{x}%
\boldsymbol{\cdot\theta})}{r^{5}},\quad\mathrm{d)}\;\; r^{2}%
(\boldsymbol{B\cdot\theta})-3(\mathbf{x\cdot B})(\mathbf{x}\boldsymbol{\cdot
\theta}). \label{a-d}%
\end{equation}
Again, the same as before, the condition that the field strength should
decrease for large distances $r$ from the charge is not sufficient to fix the
solution: more boundary conditions are needed as an effect of gauge invariance
violation by external current in the NC theory as discussed in
Sec.\ref{sec-Max}. We may discard solution d) as giving rise to (anisotropic)
electric field (linearly) growing in the remote region, but we must note that
such a solution makes an interesting option of an external field admitted by
sourceless equations of motion. The constant solution a) should be disregarded
as a pure gauge, left still unfixed after the gauge conditions used above were
imposed. (Remind that in the $U(1)$ gauge theory the Lorentz gauge imposed to
reduce the Maxwell equations for the field strengths to the Laplace equations
for potentials in a stationary problem, where fields and potentials do not
depend on time $t$, turns into the Coulomb gauge for the 3-vector potential.
However, there remains a residual gauge freedom determined by the gauge
parameter $\lambda=\lambda_{1} t+\lambda_{2}(\mathbf{x})$ with $\lambda_{1}$
being a constant and $\boldsymbol{\nabla}^{2}\lambda_{2}(\mathbf{x})=0$.
Therefore, the scalar potential $A_{0}$ remains fixed only up to this constant
$\lambda_{1}$ and up to a function $\lambda_{2}$ subject to the homogenous
Laplace equation). By linearly combining the remaining two solutions b) and c)
(\ref{a-d}) with the solution (\ref{sw24}) one can form all solutions,
satisfying boundary conditions, other than (\ref{es4}), but still, perhaps,
also physically reasonable. Let us discuss such possibilities. First, solution
b) multiplied by the constant factor $gZe$ may be added to solution
(\ref{sw24}) to exclude from it the Coulomb (rightmost in (\ref{sw24})) term.
Note, however, that the other long-range correction to the Coulomb potential
$gZe(\mathbf{x\cdot B})(\mathbf{x}\boldsymbol{\cdot\theta})/r^{3}$ still
cannot be excluded. Second, solution c) multiplied by $gZea^{2}/5$, when added
to (\ref{sw24}), leads to a solution of (\ref{sw21}), (\ref{sw22}) free of
electric quadrupole term (see eq. (\ref{quadrupole}) in he next section). The
both new solutions, as well as any other solution formed by combining b) and
c) with (\ref{sw24}), are singular in the origin, although b) does not produce
the fall down onto the center. Quite the opposite, combining b) with
(\ref{sw24}) would lead to a solution with the $1/r$ behavior of the potential
that is considered as admissible in the standard theory.

We have now to answer the question set in Subsection \ref{subsection3.2},
bearing in mind that an applicability for use cannot serve a sufficient
criterium for fixing a physical solution of the field-and-current equations of
motion in the NC theory. To finally ground why we are keeping to the choice of
(\ref{sw24}) and (\ref{es5}) as the only appropriate solution is the
following. Let us remember that the principal motivation for creating a NC
theory was the complete ultraviolet regularization by the elementary length of
everything, including the Coulomb potential itself, because many ultraviolet
troubles are already due to this weakest singularity, to say nothing of
stronger singularities in the origin that might appear in the theory.
Solutions (\ref{sw24}) and (\ref{es5}) are the only ones among other
possibilities discussed in the present Section that are totally free of any
singularities in the origin. In other words, these are the only solutions,
\textit{regularizable} by the charge size. Naturally, these should not
necessarily survive the limiting transition to a point charge, since the
latter notion is away from the NC theory. Indeed, eq. (\ref{es5}) does not.

Another, more technical, reason to stick to these solutions lies in the
validity of the approximation considered. The effective parameters used in the
expansion here are $f^{2}\theta$ and $fj\theta$. Both of them remain small on
the nonsingular solutions chosen, (\ref{es5}) and (\ref{sw24}), even when the
size $a$ is taken as its minimum $a=\sqrt{\theta}$. Namely, $f\theta
=g(Ze)^{2}$ for (\ref{es5}) and $f\theta=gZeB\theta$ for (\ref{sw24}). So only
the values of the charge $Z$ and the external magnetic field $b$ are
restricted. A use of any singular solution would lead us out the applicability
domain for sufficiently small $r$.

\section{Properties of regular solutions}

\label{sec-form}

Solutions (\ref{es5}) and (\ref{sw24}) provide stationary long-distance
corrections to the zeroth order potential (\ref{13}) induced by a static
spherical charge. These corrections may be understood as higher order
form-factors of a finite size spherical charge induced by the
noncommutativity, because they can be interpreted in terms of appearance of an
effective charge density surrounding that charge, as well as of the dipole
magnetic and quadrupole electric moments.

Irrespective of whether the external magnetic field $\mathbf{B}$ is present, a
magnetic field carried by $A^{\left(  1\right)  k}(\mathbf{x})$ proportional
to $\theta$ and independent of the external magnetic field is induced. On the
contrary, the electric field remains unchanged within the first order of
$\theta$ if $\mathbf{B=0}$, but gains first-order corrections if the external
magnetic field is present. The spherically-symmetric source does not undergo corrections.

\subsection{Magnetic dipole}

\label{subsection4.1} The leading long-distance part of the vector-potential
(\ref{es5}) behaves like that of a magnetic dipole, the static charge
(\ref{10.1}) being thus a carrier of an equivalent magnetic moment
$\boldsymbol{\mathcal{M}}$%
\begin{equation}
\boldsymbol{A}= \frac{\boldsymbol{\mathcal{M}}\times\mathbf{x}}{r^{3}},
\qquad\boldsymbol{\mathcal{M}}= \boldsymbol{\theta}~ (Ze)^{2}\frac{2g}{5a}
\label{magnmoment}%
\end{equation}
(we used the designation $\times$ for the vector product). Though the magnetic
moment grows infinitely in the limit of a point-like charge, $a\rightarrow0,$
this fact should not be thought of as a trouble, since the size of the charge
is restricted from below, $a>l$, by the elementary length $l=\sqrt{\theta}$.

Taking expression (\ref{magnmoment}) for the magnetic moment, in Ref.
\cite{AdoGitShaVas11} we studied an efficient tool of getting stronger bounds
on the noncommutativity parameter basing on the fact that in all scattering
processes leptons do not show any size at all. Once theoretical calculations
of the lepton anomalous magnetic moments, based on standard commutative
models, do not contradict their observable values within the existing
experimental and theoretical accuracy, we admitted that, at the worst, all
clearance in their values may be attributed to the effect of noncommutativity
described above. Then, as long as we relied on the existing experimental
bounds of the lepton size, we obtained that the noncommutativity parameter is
bounded by the values following already from other present-days estimates. But
admitting the point-likeness of the electron we got the hitherto strongest
bound of 10$^{4}$ TeV among the ones based on particle physics experiments.

The magnetic moment of a proton should contribute to the hyperfine splitting
of the $1S_{1/2}$-states in the hydrogen atom.
%, its order of magnitude does not depend upon a
%choice under consideration.
When calculated with the help of the outer part of (\ref{es5}), the spitting
is proportional to $(1/a)\overline{r^{-3}}$, where the bar means averaging
over $r>a$ in the $S$-state outside of the proton, with $a$ now taken as its
size. On the other hand with the outer part of solution (\ref{15}) the
corresponding contribution \cite{Stern} makes $\overline{r^{-4}}$. The two
expressions are of the same order of magnitude, because the averaging
effectively results in the substitution $r=a,$ owing to the singular character
of the averaged function and the fact that the proton size $a$ is much less
than the Bohr radius $a_{0}=\hbar^{2}/m_{\mathrm{e}}\alpha$, where
$m_{\mathrm{e}}$ is the electron mass and $\alpha=1/137$ is the fine structure
constant. So, taking into account the noncommutative magnetic moment of the
proton does not change the existing bound on the noncommutativity parameter.

There is a different context \cite{DuvHor}, wherein the noncommutativity of
coordinates is introduced associated with a charged particle spin. Then, most
naturally, it also carries a magnetic moment.

A production of magnetic field by a static electric charge - the
magneto-electric effect - was reported in Quantum Electrodynamics with a
constant and homogeneous external (magnetic plus electric) field of the most
general form \cite{ShaUso10}. The inhomogeneous magnetic field produced by a
static charge in that problem exists in an approximation linear in the charge,
when the charge is small. Contrary to that situation, in the present problem
we have found a solution of nonlinear Maxwell equations (\ref{SWeq}),
(\ref{SWin}) within the first order of $\theta$, and this solution is, for its
magnetic component, quadratic in the charge $eZ$, as it is seen from
(\ref{es5}). (The same statement holds true for the solution (\ref{15}).) The
absence of a linear-in-the-charge part of the magnetic field is in agreement
with the statement in \cite{FreGitSha11} that in NC electrodynamics without a
background field the photon polarization tensor, responsible for the linear
response, is zero in spite of the presence of the noncommutativity tensor
$\theta^{ij}$.

\subsection{Enhancement of the Coulomb law and electric quadrupole}

\label{subsection4.2} Let us now turn to the combined effects of
noncommutativity and external homogeneous magnetic field, which is the
correction (\ref{sw24}) to the electrostatic potential. It is worth noting
that this correction is linear in the charge $eZ$. This corresponds to the
fact that now that there is a homogeneous magnetic-field background the linear
response tensor is no longer trivial \cite{FreGitSha11}, although yet unable
to provide the magneto-electric effect, so the magnetic correction remains
quadratic in the charge $eZ$. As for the leading behavior of (\ref{sw24}) in
the remote region $r\gg a$, it follows the Coulomb law $\sim1/r$. When united
with (\ref{13}), it gives the anisotropic, NC-corrected Coulomb potential
\begin{equation}
A_{\mathrm{Coulomb}}^{0}\left(  \mathbf{x}\right)  =\left(  \frac{Ze}%
{r}\right)  \left(  1-g\left\{  \frac{1}{r^{2}}\left(  \mathbf{x}%
\cdot\mathbf{B}\right)  \left(  \mathbf{x}\cdot\boldsymbol{\theta}\right)
+\left(  \mathbf{B}\cdot\boldsymbol{\theta}\right)  \right\}  \right)
\,,\quad r\gg a\,. \label{Coulomb}%
\end{equation}
The correction may be attributed to the $1/r^{3}$ behavior of the right-hand
side of eq. (\ref{sw22}) - the \textquotedblleft dark charge
density\textquotedblright\ distribution. (We use this term in analogy with the
notion of the dark matter introduced to resume the responsibility for the
observed gravitational field deviation from the Newtonian law). In the special
case where the external field is oriented in parallel (antiparallel) to the
noncommutativity vector, $\mathbf{B}\parallel\pm\boldsymbol{\theta},$ the
overall multiplier of the standard Coulomb potential $eZ/r$ in eq.
(\ref{Coulomb}) becomes $1\mp g|B||\theta|(\cos^{2}\vartheta+1)$, where
$\vartheta$ is the angle between $\mathbf{x}$ and $\mathbf{B.}$ For the
antiparallel configuration (lower sign) with positive charge $g>0$ (as well as
for parallel configuration with negative charge) the correction to the unity
in this formula is positive for every direction of the radius-vector. The
latter result allows us to estimate the maximum value of the long-range NC
correction to the Coulomb field, which reads in either of the above cases
\begin{equation}
A_{\mathrm{LR}}^{0}\left(  r\right)  =\frac{Ze}{r}\left[  1+gB\theta)\right]
\simeq\frac{Ze}{r}\left(  1+\delta\right)  \,. \label{LR}%
\end{equation}
Assuming a magnetic field of magnitude $10$ Tesla (which is a very strong
laboratory field) and the NC parameter of $(100\mathrm{TeV})^{-2}$ (which does
not satisfy the strongest bounds on the NC scale, see e.g.
\cite{AdoGitShaVas11}), we obtain $\delta=6\cdot10^{-26}$ which is far beyond
possibilities of any experimental verification. Even for the magnetic field on
the surface of Soft Gamma Repeater which reaches $10^{11}\mathrm{T}$
\cite{SGR} and the same $\theta$ as above the correction $\delta
=6\cdot10^{-16}$ is very small.

On the other hand, for the same special cases when the magnetic field
$\mathbf{B}$ and the vector $\boldsymbol{\theta}$ are parallel or
antiparallel, the angular dependence of solution (\ref{sw24}) leads to
splitting between levels in an atom with different angular momentum
projections onto the common direction of these vectors to compete with the
Zeeman splitting (at a much lower level, of course).

We saw that Coulomb field produced by a charge in external magnetic field far
away from the former is enhanced as compared to Maxwell electrodynamics. This
unprecedented property is absent from QED, where the linear in the charge
correction to (the long-range part of) the Coulomb potential only makes it
anisotropic without enhancing it \cite{ShaUs07, SadJal}: the potential
decreases as $1/r$ along the magnetic field following the same Coulomb law as
in empty space, and it decreases along any other direction $\vartheta\neq0$
also following the Coulomb law, but taken with the coefficient $(\cos
^{2}\vartheta+\beta\sin^{2}\vartheta)^{-1/2}$ smaller than unity. Here $\beta=
(1+\frac{\alpha}{3\pi}\frac{eB}{m_{e}^{2}})^{1/2}>1$.

Note that the leading (Coulomb) part (\ref{Coulomb}) survives the limiting
transition $a\to0$ to the point-like charge. Also, when there is no external
magnetic field, the (cubic in the charge) correction \cite{BerLifPit} to
electrostatic potential in QED does not affect its long-ranged Coulomb part.

The long-distance next-to-leading part in $A_{II}^{(1)0}$ corresponds to an
equivalent electric quadrupole moment $D_{ij}$%
\begin{equation}
A^{0}=\frac{D_{ij}x_{i}x_{j}}{r^{5}}, \qquad D_{ij}=2gZea^{2}(3B_{i}\theta
_{j}-\delta_{ij}(\boldsymbol{B\cdot\theta})) \label{quadrupole}%
\end{equation}
that may be attributed to the finite-size charge (\ref{10.1}), although it is
spherically symmetric. The NC quadrupole moment vanishes in the limit $a\to0$.

There is no electric-dipole part in $A_{\mathrm{II}}^{(1)}$. In this respect
the situation is similar to QED, where the post-Coulomb long-range tail in the
potential produced by a spherically symmetric charge in a magnetic field does
not contain the dipole $1/r^{2}$ term either, but starts, according to
\cite{SadJal}, with $1/r^{3}$, the same as (\ref{sw24}).

\subsection{Powers of the charge and generalized Furry theorem}

\label{subsection4.3} A remark is in order, which provides one with a tool -
that may be referred to as a generalized Furry theorem - to judge, prior to
calculations, about the powers of the charge $eZ$ on general grounds within
and beyond the leading approximation of the first order of $\theta$. As long
as the space-space NC theory conserves parity, the vector $\boldsymbol{\theta
}$ is a pseudovector, the same is a magnetic field. Referring to the QFT
language, we can say, that this implies that the total number of legs in a
many-photon diagram, characteristic of the nonlinear theory under study,
connected to a magnetic field and to the \textquotedblleft
field\textquotedblright\ $\boldsymbol{\theta}$ must be even. Due to the Furry
theorem that states that the overall number of photon legs should be even, we
conclude that the number of photon legs connected to electrostatic field
should be even separately. So, the magnetic field, produced by the static
charge in the approximation, linear in $\theta$, irrespective of whether the
external magnetic field is present or not, must be even in the charge: the
corresponding diagram contains one $\theta$-leg, an even number or none of
legs joining it to the external magnetic field ($B$-legs), one leg
corresponding to the produced magnetic field ($b$-leg) and even number of legs
joining to the external charge ($Z$-legs). This is in accordance with the
result (\ref{es5}), wherein the latter number is two. On the other hand the
electric field cannot undergo a correction linear in $\theta$ if external
magnetic field is absent. In that case an impossible configuration with one
$\theta$-leg would be required. Such correction might be only of even order in
$\theta$, that fell beyond our consideration. The situation changes when
external magnetic field is turned on. Now there is an admissible configuration
of one $\theta$-leg and an odd number of $B$-legs, so the first-order
correction to the electric field should include one leg going to the produced
electric field ($e$-leg) and an odd number of $Z$-legs (hence, odd powers of
the charge $Z$) to keep the total numbers of legs connecting with the
electrostatic field even. Out of these odd powers of the charge we got only
one, because the keeping of only the first power of $\theta$ used when
deriving the field equations (\ref{SWeq}), (\ref{SWin}) has essentially
reduced the otherwise unlimited extent of nonlinearity in the field.

When applied to the standard QED with no NC parameter, the generalized Furry
theorem explains why the electric field produced by a static charge, besides
being proportional to the value of the latter, has also contributions odd in
it (the cubic contribution \cite{BerLifPit} was mentioned above). It also
predicts the existence in QED of a magnetic field, quadratic in the static
charge value, produced by such charge, when it placed into external strong
magnetic field -- yet another manifestation of the magneto-electric effect.

%%%%%%%%%%%%

\section{Ambiguities in the SW map}

\label{sec-amb}

It is very well known that the SW map is not uniquely defined. There are
additional terms which can be interpreted as redefinition of the gauge fields
\cite{AsaKis99}. Such terms have been discussed in the context of
renormalization of the noncommutative Maxwell theory
\cite{BicGriGroPopSchWul01,FruGriMorPopSch02}, noncommutative Dirac fields
coupled with the Yang-Mills \cite{Wulkenhaar02} and of the noncommutative
chiral electrodynamics \cite{BirLatRadTra10}. In the case of noncommutative
$U\left(  1\right)  $ gauge theories, it was showed (e.g.
\cite{BicGriGroPopSchWul01}) that the SW map to the potentials admits, in
first order in $\theta$, the following extension,%
\begin{align}
&  \check{A}_{\mu}\left(  { x}\right)  =A_{\mu}\left(  { x}\right)  +\frac
{g}{2}\theta^{\alpha\beta}A_{\alpha}\left(  { x}\right)  \left[
\partial_{\beta}A_{\mu}\left(  { x}\right)  +f_{\beta\mu}\left(  {x}\right)
\right]  +\mathbb{A}_{\mu}\left(  { x}\right)  \,,\nonumber\\
&  \mathbb{A}_{\mu}\left(  { x}\right)  =g\kappa_{1}\theta_{\mu\alpha}%
\partial_{\beta}f^{\alpha\beta}\left(  { x}\right)  \,, \qquad\kappa
_{1}=\mathrm{const}\,, \label{sw6.0}%
\end{align}
that keeps the Euler-Lagrange equations the same as they are in the pure
noncommutative $U\left(  1\right)  $ theory with the action $\check
{S}_{\mathrm{A}}$ (\ref{SAjA}). To see this it is sufficient to form
$f^{\nu\mu}$ out of $\mathbb{A}_{\mu}\left(  { x}\right)  $ and make sure that
it gives vanishing contribution into the first term of eq. (\ref{SWeq}).
Therefore $\kappa_{1}$ does not appear in equations of motion to the
first-order accuracy. Once $\mathbb{A}_{\mu}\left(  { x}\right)  $ satisfies
the homogeneous part of the equation of motion, the transformation
(\ref{sw6.0}) reduces to adding such solution of the source-free equations to
any other solution of eq. (\ref{SWeq}).

A very similar ambiguity in the SW equations for currents (\ref{sw4.1}) was
observed in \cite{Ban}. One can add two extra terms $\mathbb{J}^{\mu}\left(
x\right)  $ to the solution (\ref{sw4.2}):
\begin{align}
\check{j}^{\mu}\left(  x\right)   &  =j^{\mu}\left(  x\right)  +g\theta
^{\alpha\beta}A_{\alpha}\left(  x\right)  \partial_{\beta}j^{\mu}\left(
x\right)  +\mathbb{J}^{\mu}\left(  x\right)  \,,\nonumber\\
\mathbb{J}^{\mu}\left(  x\right)   &  =g\left(  \kappa_{2}\theta^{\alpha\beta
}f_{\alpha\beta}j^{\mu}+\kappa_{3}\theta^{\mu\alpha}f_{\alpha\beta}j^{\beta
}\right)  \,,\qquad\kappa_{2},\kappa_{3} =\mathrm{const} \,. \label{sw6.1}%
\end{align}
It is easy to check that the current (\ref{sw6.1})\ satisfies (\ref{sw4.1})
for arbitrary values of $\kappa_{2}$ and $\kappa_{3}$ already because
$\mathbb{J}^{\mu}\left(  x\right)  $ does not undergo gauge transformation.

Let us check how the ambiguities described above influence the solutions of NC
Maxwell equations. At the zeroth order in $\theta$, the equations remain the
same, see (\ref{Max0}). At the first order,
\begin{equation}
\mathbb{A}_{\mu}^{(1)} = g\kappa_{1}\theta_{\mu\alpha} \partial_{\beta
}f^{(0)\alpha\beta}=-\frac{4\pi}c g\kappa_{1}\theta_{\mu\alpha} j^{(0)\alpha
}\,.
\end{equation}
Obviously, $\mathbb{A}_{\mu}^{(1)}$ vanishes for a static charge distribution
in the case of a space-space noncommutativity. So, in our special problem the
ambiguity $\mathbb{A}_{\mu}^{(1)}$ does not work at all. The Maxwell equation
and the compatibility conditions at the order $\theta$ read
\begin{align}
&  \partial_{\nu} f^{(1)\nu\mu}-g\theta^{\alpha\beta} \left(  \partial_{\nu
}(f_{\alpha}^{(0)\nu}f_{\beta}^{(0)\mu}) -f_{\nu\alpha}^{(0)}\partial_{\beta
}f^{(0)\nu\mu} \right)  =\frac{4\pi}c (j^{(1)\mu}+\mathbb{J}^{(1)\mu
})\,,\label{SWeq-k}\\
&  \partial_{\mu}(j^{(1)\mu}+\mathbb{J}^{(1)\mu})+g\theta^{\alpha\beta}
f_{\mu\alpha}^{(0)}\partial_{\beta}j^{(0)\mu}=0\,, \label{SWin-k}%
\end{align}
where $\mathbb{J}^{(1)}$ is a given function of the zeroth order
electromagnetic potential and the current
\begin{equation}
\mathbb{J}^{(1)\mu} =g\left(  \kappa_{2}\theta^{\alpha\beta}f_{\alpha\beta
}^{(0)}j^{(0)\mu}+\kappa_{3}\theta^{\mu\alpha}f_{\alpha\beta}^{(0)}%
j^{(0)\beta}\right)  \,. \label{kkJ}%
\end{equation}

The equations (\ref{SWeq-k}) and (\ref{SWin-k}) do not depend on $\kappa_{1}$,
while $\kappa_{2}$ and $\kappa_{3}$ enter both equations only through the
combination $j^{(1)\mu}+\mathbb{J}^{(1)\mu}$. Besides, this combination is
defined by exactly the same equation as in the case $\kappa_{2}=\kappa_{3}=0$,
see eq.\ (\ref{SWin1}). Therefore, the whole ambiguity in the first order
corrections to the electromagnetic potentials and to the \textquotedblleft
acting\textquotedblright\ current $j^{(1)\mu}+\mathbb{J}^{(1)\mu}$ is no
ampler than the natural arbitrariness of adding homogeneous solutions to
solutions of equations (\ref{SWin1}) and (\ref{SWin-k}). (This ambiguity has
been already discussed above, see the paragraph below eq. (\ref{SWin1}).) On
the other hand separate parts in the combination $j^{(1)\mu}+\mathbb{J}%
^{(1)\mu}$ remain ambiguous, the physical results being independent of any
separation of the acting current into parts. The same conclusions can be drawn
dealing directly with eqs. (\ref{SWeq}), (\ref{SWin}) without appealing to the
special case of space-space noncommutativity and stationarity.

\section{Conclusions}

In this paper we have studied how one can introduce external currents
(sources) in the classical NC Maxwell theory without violating the gauge
covariance. We started with a $U(1)_{\star}$ gauge theory and found that this
can be selfconsistently done at the level of the equations of motion. Note,
that in this case the currents transform under the gauge transformations, as
well as the field strength. Further, we argued that to facilitate the
comparison with predictions of commutative electrodynamics one needs the
fields with the same gauge-transformation properties as in the commutative
case. A transition to such fields is done by means of the SW map, and we
extended this map to include the currents. Again, a consistent result is
obtained if one works at the level of the Maxwell equations rather that at the
level of the action. We wrote weakly nonlinear anisotropic equations of
motion, wherein the fields and the currents are involved, that are valid up to
the first power of the NC parameter $\theta$. Although these equations contain
potentials along with the field strengths they are gauge covariant in the
sense that the gauge transformed potentials satisfy the same equations,
moreover the potentials can be on-shell eliminated from the equations, i.e.
the equations with the potential-containing terms omitted have common
solutions with the primary equations. For the case of space-space
noncommutativity we considered an example, where the external source is a
homogeneously charged sphere of finite radius and solved the equations of
motion in the presence of an external constant and homogeneous magnetic field.
No first-order in $\theta$ correction to the source appears in the spherically
symmetric problem under consideration. To select solutions we impose the
boundary conditions that require that these be finite in the point where the
source is centered. The magnetic solution fixed in this way \textit{does not}
admit the limiting transition to a point source, which is an admissible
property, since in the noncommutative theory a size of a physical body cannot
be smaller than the elementary length.

We studied the contents of the solutions obtained. We found angle-dependent
correction to the electric field produced by a static charge that implies an
enhancement of the Coulomb law in the remote region of space in the presence
of a constant magnetic field -- a remarkable macroscopic consequence of the
microscopic elementary length inherent to the NC electrodynamics under study.
We found also that the next-to-leading behavior of the electric field far from
the charge in the presence of external magnetic field is the one of an
electric quadrupole. We noted an uncustomary possibility contained in item d)
of eq.(\ref{a-d}) that an anisotropic electric field linearly growing with the
distance from the charge and nonsingular in the origin satisfies the field
equations and can be therefore thought of as another admissible external field
added to the constant magnetic field already present or a constant electric
field. Irrespective of whether the external magnetic field is present or not,
according to the chosen solution, the static charge, apart of giving rise - as
usual - to an electrostatic field, also behaves itself as a magnetic dipole,
with the magnetic moment depending on its size and proportional to the second
power of the charge.

Finally, we studied the ambiguities in the definition of the SW map in the
presence of currents, and found that at the first order in $\theta$ this is
precisely the ambiguity of adding a homogeneous solution in the current
conservation equation.

\section*{Acknowledgements}

This work was supported by FAPESP (T.C.A., D.M.G.\ and D.V.V.), CNPq
(D.M.G.\ and D.V.V.) and by the Russian Foundation for Basic Research, Project
No. 11-02-00685-a (A.E.S.).

\end{document}